\documentclass[12pt]{article}

\usepackage[T1]{fontenc}
\usepackage[utf8]{inputenc}
\usepackage[margin=1in]{geometry}
\usepackage{amsmath,amssymb}
\usepackage{graphicx}
\usepackage{booktabs}
\usepackage{multirow}
\usepackage{newtxtext,newtxmath}
\usepackage[font=small,labelfont=bf]{caption}
\usepackage{microtype}

\usepackage{natbib}
\bibpunct{(}{)}{;}{a}{}{,}
\usepackage[hidelinks,breaklinks]{hyperref}
\newcommand{\Dn}{\Delta n}                     
\newcommand{\mut}{\mu}                         
\newcommand{\Str}{\mathcal{S}}                 

\newcommand{\Ss}{S_{1}}
\newcommand{\Sss}{S_{2}}

\begin{document}

\title{Derivation of OH sum rules without Rayleigh--Jeans approximation:
a powerful tool for measuring the variation of fundamental physical constants}

\author{Renzhi Su$^{\ast}$, Minfeng Gu\\[6pt]
\small Shanghai Astronomical Observatory, Chinese Academy of Sciences,\\
80 Nandan Road, Shanghai 200030, China\\
\textit{Email:} \texttt{rzsu.astro@gmail.com}}

\date{\today}

\maketitle

\begin{abstract}
In this paper, we show that the optical depths of any OH $\Lambda$-doublet quartet obey one exact linear relation in every velocity channel. No approximation is needed. The only exception is the $^{2}\Pi_{1/2}$ $J{=}1/2$ level, which has three transitions only, because its $F{=}0{\leftrightarrow}0$ line is strictly forbidden. The relation is $\tau_{S_1}(v)/\mu_{S_1}^{2}+\tau_{S_2}(v)/\mu_{S_2}^{2}=\tau_{M_1}(v)/(a\mu_{M_1}^{2})+\tau_{M_2}(v)/(b\mu_{M_2}^{2})$, where $a=2J^{2}+J-1$, $b=2J^{2}+3J$, and the $\mu_i^{2}$ are reduced transition dipole moments derived from the catalogued Einstein $A$ coefficients. It holds equally for gas in local thermodynamic equilibrium (LTE), for non-LTE gas, for masers and for conjugate gas. We suggest that it is a powerful tool for measuring the variation of fundamental physical constants.
\end{abstract}

\section{Introduction}

\begin{figure}[htbp]
\centering
\includegraphics[width=0.75\linewidth]{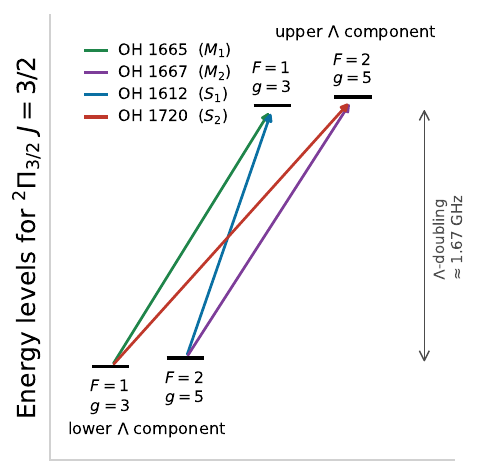}
\caption{The $^{2}\Pi_{3/2}$ $J{=}3/2$ ground-state quartet of OH. $g=2F+1$ is the statistical weight. Coloured arrows mark the four lines: main lines $M_1$ (1665~MHz) and $M_2$ (1667~MHz), satellite lines $S_1$ (1612~MHz) and $S_2$ (1720~MHz).}
\label{fig:levels}
\end{figure}

The hydroxyl radical OH was the first molecule detected in the interstellar medium at radio wavelengths \citep{weinreb1963}, and it remains one of the most useful molecules in radio astronomy. The coupling between the electronic orbital angular momentum and the molecular rotation splits every rotational level into a pair of parity states, the well-known $\Lambda$-doubling \citep{dousmanis1955}. The proton spin $I{=}1/2$ splits each parity state further into hyperfine sublevels. The lowest rotational state, $^{2}\Pi_{3/2}$ $J{=}3/2$, therefore has four sublevels, connected by four radio lines near 18~cm (Fig.~\ref{fig:levels}).  These lines have been seen in emission from diffuse interstellar gas, in absorption toward bright continuum sources, and as masers around evolved stars and in star-forming regions \citep[e.g.][]{robinson1967,elitzur1992,lo2005,dawson2022}.

OH lines are also a useful tool for measuring fundamental constants. $\Lambda$-doubling and the hyperfine splitting depend, in different ways, on the fine-structure constant $\alpha$, the proton-to-electron mass ratio $m_{p}/m_{e}$, and the proton $g$-factor $g_{p}$, so comparing the four lines tests whether these constants change over cosmic time \citep{uzan2025,chengalur2003}. Astronomical observations have already placed tight constraints on the evolution of the constants \citep[e.g.][]{darling2004,kanekar2005,gupta2018,kanekar2018,su2025a,su2025b,su2026}.

Reference~\cite{su2026a} (hereafter SG26) recently combined the four OH 18~cm optical depths into the sum rule $\tau_{1612}+\tau_{1720}=\tau_{1665}/5+\tau_{1667}/9$, and used it to measure the variation of fundamental physical constants. Their derivation, however, requires the excitation temperatures of the lines and relies on the Rayleigh--Jeans approximation, $kT_{\rm ex} \gg h\nu$.

In this paper we derive the relation again, but without the Rayleigh--Jeans approximation. All relations presented here hold for arbitrary excitation --- local thermodynamic equilibrium (LTE), non-LTE, inversion, or enhanced absorption --- and for optical depths of any sign and size, provided that the four lines are observed along a common line of sight, that they share a common profile in velocity space, and that the optical depth is defined as in Eq.~(\ref{eq:tau}).

\section{Derivation of the sum rule}
\label{sec:derive}

\subsection{From the optical depth to the level populations}
\label{sec:tau}

The optical depth of a line $i$ in a velocity channel $v$ is
\begin{equation}
\tau_i(v)
=\frac{c^{3}}{8\pi\nu_i^{3}}\,A_i\,\frac{g_{u,i}}{g_{l,i}}\,\Dn_i\,\phi_i(v) ,
\label{eq:tau}
\end{equation}
where $\nu_i$ is the rest frequency, $A_i$ the Einstein coefficient, $g_{u,i}$ and $g_{l,i}$ the statistical weights of the upper and lower levels, $\phi_i(v)$ the line profile, normalized in velocity so that $\int\phi_i\,{\rm d}v=1$, and
\begin{equation}
\Dn \equiv n_l-\frac{g_l}{g_u}\,n_u
\label{eq:deficit}
\end{equation}
the weighted difference of the two level column densities $n_l$, $n_u$.

The Einstein coefficient can be written as
\begin{equation}
A_i=\frac{16\pi^{3}\nu_i^{3}}{3\varepsilon_0 h c^{3}}\,
\frac{|\langle l_i\|\hat{\mu}\|u_i\rangle|^{2}}{g_{u,i}} ,
\label{eq:Afact}
\end{equation}
where $|\langle l_i\|\hat{\mu}\|u_i\rangle|^{2} $ is the squared reduced matrix element, which can be factored into dimensionless hyperfine strength factor $\Str_i$ and reduced transition dipole moment $\mut_i$,
\begin{equation}
|\langle l_i\|\hat{\mu}\|u_i\rangle|^{2}=\Str_i\,\mut_i^{2} .
\label{eq:factor}
\end{equation}
Equivalently, once $A_i$ has been obtained, Eq.~(\ref{eq:factor}) defines the reduced transition dipole moment line by line,
\begin{equation}
\mut_i^{2}\equiv\frac{3\varepsilon_0 h c^{3}}{16\pi^{3}\nu_i^{3}}\,
\frac{A_i\,g_{u,i}}{\Str_i} ,
\label{eq:mudef}
\end{equation}
which is how the moments in Table~\ref{tab:bands} are computed. Taking Eqs.~(\ref{eq:Afact}) and (\ref{eq:factor}) into Eq.~(\ref{eq:tau}) gives
\begin{equation}
\frac{\tau_i(v)}{\Str_i\,\mut_i^{2}}
=\frac{2\pi^{2}}{3\varepsilon_0 h}\,
\frac{\Dn_i\,\phi_i(v)}{g_{l,i}} .
\label{eq:key}
\end{equation}
Every line-specific quantity has been moved to the left-hand side; the right-hand side has only the populations and the profile.

\subsection{The quartet and the population identity}
\label{sec:quartet}

Now consider a rotational level $J$ of OH. $\Lambda$-doubling splits it into two parity states $+$ and $-$, and the proton spin $I{=}1/2$ splits each of them into $F_1=J-\tfrac12$ and $F_2=J+\tfrac12$, with statistical weights $g_1=2J$ and $g_2=2J+2$. Four lines connect the two parities; the main lines $M_1$ and $M_2$ with $\Delta F=0$, and the satellite lines $S_1$ and $S_2$ with $\Delta F=\pm1$. Figure~\ref{fig:levels} shows the ground-state quartet, $^{2}\Pi_{3/2}$ $J{=}3/2$.

Denoting the populations of the four levels by $n_1^-$, $n_2^-$, $n_1^+$ and $n_2^+$, the four weighted differences of the quartet are
\begin{align}
\Dn_{\Ss}&=n_2^--\frac{g_2}{g_1}\,n_1^+, &
\Dn_{M_1}&=n_1^--n_1^+, \nonumber\\
\Dn_{\Sss}&=n_1^--\frac{g_1}{g_2}\,n_2^+, &
\Dn_{M_2}&=n_2^--n_2^+ ,
\label{eq:deficits}
\end{align}
from which we obtain:
\begin{equation}
\frac{\Dn_{M_1}}{g_1}+\frac{\Dn_{M_2}}{g_2}
=\frac{\Dn_{\Ss}}{g_2}+\frac{\Dn_{\Sss}}{g_1} .
\label{eq:identity}
\end{equation}
Note that no assumption has been made. Equation~(\ref{eq:identity}) is pure algebra, and therefore holds whether the gas is in LTE, out of LTE, under strong far-infrared pumping, or in a saturated maser.

\subsection{The line strengths: the coefficients $a$ and $b$}\label{sec:ab}

For a level $J$ coupled to a nuclear spin $I$, angular-momentum algebra gives the relative strength of the hyperfine component $F{\leftrightarrow}F'$ as
\begin{equation}
\Str_{FF'}=(2F+1)(2F'+1)
\begin{Bmatrix} J & F' & I \\ F & J & 1 \end{Bmatrix}^{2}
\label{eq:sixj}
\end{equation}
\citep{varshalovich1988}. With $F_1=J-\tfrac12$, $F_2=J+\tfrac12$, and  $I{=}1/2$, the relative line strength of the four lines is
\begin{align}
\Str_{\Ss}&=\Str_{\Sss}=\frac{2}{(2J+1)^{2}} , \nonumber\\
\Str_{M_1}&=\frac{2\,(2J^{2}+J-1)}{(2J+1)^{2}} , \nonumber\\
\Str_{M_2}&=\frac{2\,(2J^{2}+3J)}{(2J+1)^{2}} .
\label{eq:sclosed}
\end{align}
Normalizing to the satellite lines gives
\begin{equation}
a\equiv\frac{\Str_{M_1}}{\Str_{S}}=2J^{2}+J-1,\qquad
b\equiv\frac{\Str_{M_2}}{\Str_{S}}=2J^{2}+3J ,
\label{eq:ab}
\end{equation}
where $a$ and $b$ are the main-line strengths relative to the satellite lines. Table~\ref{tab:bands} lists $a$ and $b$ for the $J$ values that occur in the OH radio bands. Note that $J{=}1/2$ gives $a=0$, because the $^{2}\Pi_{1/2},J{=}1/2,F{=}0{\leftrightarrow}0$ main line is strictly forbidden.

\begin{table}[htbp]
\centering
\caption{The OH $\Lambda$-doublet radio bands. Rest frequencies and Einstein $A$ coefficients are taken from the LAMDA compilation \citep{schoier2005}. $S_1$ and $S_2$ are the satellite lines, $M_1$ and $M_2$ the main lines. $\Str$ is the relative hyperfine line-strength factor obtained from Eq.~(\ref{eq:sclosed}) and $\mut_i$ the reduced transition dipole moment obtained from Eq.~(\ref{eq:mudef}). $a=\Str(M_1)/\Str(S)$ and $b=\Str(M_2)/\Str(S)$ are the relative main line strength to satellite lines. The 4.7~GHz ladder has only three lines; its $F{=}0{\leftrightarrow}0$ main line $M_1$ is strictly forbidden ($a=0$), so no sum rule exists there; its three lines are listed for completeness.}
\label{tab:bands}
\small
\setlength{\tabcolsep}{4pt}
\begin{tabular}{lllccc}
\toprule
Band & ladder & line & $\nu$ [MHz] & $A$ [s$^{-1}$] & $\Str$ \quad $\mut_i$ [D] \\
\midrule
\multirow{5}{*}{18~cm} & \multirow{5}{*}{$^{2}\Pi_{3/2}\;J{=}3/2$}
 & $S_1$ (1612) & 1612.231 & $1.283\times10^{-11}$ & $1/8$ \quad 2.51251 \\
 & & $M_1$  (1665) & 1665.402 & $7.070\times10^{-11}$ & $5/8$ \quad 2.51236 \\
 & & $M_2$  (1667) & 1667.359 & $7.663\times10^{-11}$ & $9/8$ \quad 2.51243 \\
 & & $S_2$ (1720) & 1720.530 & $9.355\times10^{-12}$ & $1/8$ \quad 2.51240 \\
 & & \multicolumn{4}{l}{\quad $a=5$,\quad $b=9$} \\
\midrule
\multirow{5}{*}{6.0~GHz} & \multirow{5}{*}{$^{2}\Pi_{3/2}\;J{=}5/2$}
 & $S_1$ & 6016.746 & $1.081\times10^{-10}$ & $1/18$ \quad 1.95894 \\
 & & $M_1$  & 6030.748 & $1.525\times10^{-9}$  & $7/9$ \quad 1.95959 \\
 & & $M_2$  & 6035.093 & $1.559\times10^{-9}$  & $10/9$ \quad 1.95928 \\
 & & $S_2$ & 6049.084 & $7.850\times10^{-11}$ & $1/18$ \quad 1.95937 \\
 & & \multicolumn{4}{l}{\quad $a=14$,\quad $b=20$} \\
\midrule
\multirow{5}{*}{13.4~GHz} & \multirow{5}{*}{$^{2}\Pi_{3/2}\;J{=}7/2$}
 & $S_1$ & 13442.114 & $3.381\times10^{-10}$ & $1/32$ \quad 1.63672 \\
 & & $M_1$  & 13434.637 & $9.115\times10^{-9}$  & $27/32$ \quad 1.63686 \\
 & & $M_2$  & 13441.417 & $9.203\times10^{-9}$  & $35/32$ \quad 1.63678 \\
 & & $S_2$ & 13433.930 & $2.625\times10^{-10}$ & $1/32$ \quad 1.63677 \\
 & & \multicolumn{4}{l}{\quad $a=27$,\quad $b=35$} \\
\midrule
\multirow{4}{*}{4.7~GHz} & \multirow{4}{*}{$^{2}\Pi_{1/2}\;J{=}1/2$}
 & $S_1$ & 4660.242 & $1.076\times10^{-9}$ & $1/2$ \quad 1.35157 \\
 & & $M_2$  & 4750.656 & $7.597\times10^{-10}$ & $1$ \quad 1.35139 \\
 & & $S_2$ & 4765.562 & $3.835\times10^{-10}$ & $1/2$ \quad 1.35150 \\
 & & \multicolumn{4}{l}{\quad $a=0$,\quad $b=2$ ($M_1$ forbidden; three lines only; no sum rule)} \\
\midrule
\multirow{5}{*}{7.8~GHz} & \multirow{5}{*}{$^{2}\Pi_{1/2}\;J{=}3/2$}
 & $S_1$ & 7749.909 & $1.855\times10^{-10}$ & $1/8$ \quad 0.90649 \\
 & & $M_1$  & 7761.747 & $9.320\times10^{-10}$ & $5/8$ \quad 0.90661 \\
 & & $M_2$  & 7820.125 & $1.029\times10^{-9}$  & $9/8$ \quad 0.90641 \\
 & & $S_2$ & 7831.962 & $1.149\times10^{-10}$ & $1/8$ \quad 0.90660 \\
 & & \multicolumn{4}{l}{\quad $a=5$,\quad $b=9$} \\
\midrule
\multirow{5}{*}{8.1~GHz} & \multirow{5}{*}{$^{2}\Pi_{1/2}\;J{=}5/2$}
 & $S_1$ & 8118.051 & $4.230\times10^{-11}$ & $1/18$ \quad 0.78189 \\
 & & $M_1$  & 8135.870 & $5.962\times10^{-10}$ & $7/9$ \quad 0.78195 \\
 & & $M_2$  & 8189.587 & $6.204\times10^{-10}$  & $10/9$ \quad 0.78189 \\
 & & $S_2$ & 8207.402 & $3.122\times10^{-11}$ & $1/18$ \quad 0.78185 \\
 & & \multicolumn{4}{l}{\quad $a=14$,\quad $b=20$} \\
\bottomrule
\end{tabular}
\end{table}

\subsection{The exact sum rule}
\label{sec:rule}
Combining Eq.~(\ref{eq:key}), Eq.~(\ref{eq:identity}) and Eq.~(\ref{eq:ab}), we obtain
\begin{equation}
\frac{\tau_{\Ss}(v)}{\mut_{\Ss}^{2}}
+\frac{\tau_{\Sss}(v)}{\mut_{\Sss}^{2}}
=\frac{\tau_{M_1}(v)}{a\,\mut_{M_1}^{2}}
+\frac{\tau_{M_2}(v)}{b\,\mut_{M_2}^{2}} ,
\label{eq:rule}
\end{equation}
which holds in every velocity channel $v$ and for arbitrary level populations. This is the exact sum rule, and it follows from the level structure alone. Its numerical accuracy is inherited from the catalogued Einstein coefficients through Eq.~(\ref{eq:mudef}).

If the four reduced transition dipole moments are treated as equal, Eq.~(\ref{eq:rule}) reduces to
\begin{equation}
\tau_{\Ss}(v)+\tau_{\Sss}(v)
=\frac{\tau_{M_1}(v)}{a}+\frac{\tau_{M_2}(v)}{b} .
\label{eq:rulesimple}
\end{equation}

For the OH 18~cm quartet, $a=5$ and $b=9$, so Eq.~(\ref{eq:rulesimple}) becomes
\begin{equation}
\tau_{1612}(v)+\tau_{1720}(v)
=\frac{1}{5}\,\tau_{1665}(v)+\frac{1}{9}\,\tau_{1667}(v) ,
\label{eq:rule18cm}
\end{equation}
which is the form quoted by SG26.

The key properties of Eq.~(\ref{eq:rule}) are worth stressing. Its only inputs are the level structure of the molecule and the definitions Eq.~(\ref{eq:tau}) and Eq.~(\ref{eq:deficit}); no excitation temperature appears anywhere, and no Rayleigh--Jeans expansion is invoked, so the rule applies in the same way to absorbing gas, to thermally emitting gas, to masers and to conjugate gas.

\begin{figure}[htbp]
\centering
\includegraphics[width=0.86\linewidth]{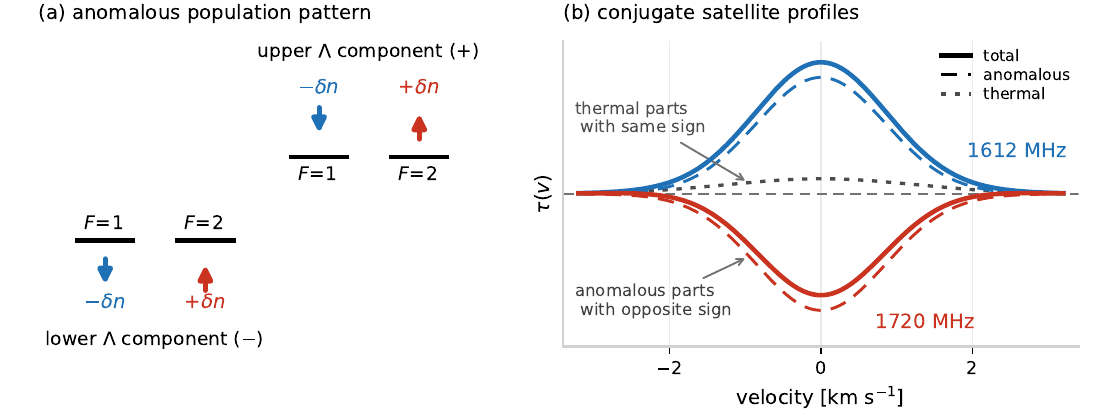}
\caption{Conjugate excitation of the 18~cm satellite lines. (a) Far-infrared pumping redistributes the ground-state populations; each hyperfine pair $F{=}1$ and $F{=}2$ is shifted by the same amount $\delta n$ in both $\Lambda$ components (arrows; here $F{=}1$ is depleted and $F{=}2$ enhanced). The main-line population differences are unchanged, while the satellite differences obey $3\,\Dn_{\Ss}^{\rm an}+5\,\Dn_{\Sss}^{\rm an}=0 $. (b) The resulting optical-depth profiles (solid curves) are the sum of a mirrored anomalous part (dashed) and a same-sign thermal part (dotted, common to both lines), so the total profiles are not exactly antisymmetric. The figure illustrates the idealised symmetric pumping discussed in the text; the two thermal components are drawn as identical curves for clarity, which is a schematic simplification.}
\label{fig:conjugate}
\end{figure}

\section{Optical depths of the conjugate 18~cm satellite lines}
\label{sec:conjugate}
Here we consider the conjugate OH 18~cm satellite lines, which are frequently seen in astronomical observations. The OH 1612 and 1720~MHz lines are called conjugate when one of them appears in absorption and the other in emission. Conjugate lines have been found from the diffuse interstellar medium of the Milky Way out to distant absorbers \citep[e.g.][]{darling2004,kanekar2018,combes2021,dawson2022,combes2023,su2026}, and their origin is usually far-infrared pumping. Absorption of 119~$\mu$m photons carries molecules from the ground ladder to $^{2}\Pi_{3/2}$ $J{=}5/2$, while 79~$\mu$m photons connect the ground ladder to the $^{2}\Pi_{1/2}$ states \citep{elitzur1992}.

Pumping shifts the populations of the two members of each pair by the same amount, $\delta n_1^-=\delta n_1^+$ and $\delta n_2^-=\delta n_2^+$, as shown in Fig.~\ref{fig:conjugate}a. This is the case for continuum far-infrared pumping. Taking these shifts into Eq.~(\ref{eq:deficits}) leaves the main-line differences unchanged and ties the satellite differences together,
\begin{equation}
3\,\Dn_{\Ss}^{\rm an}+5\,\Dn_{\Sss}^{\rm an}=0 ,
\label{eq:fmode}
\end{equation}
where $\Dn^{\rm an}$ denotes the anomalous weighted population difference produced by the pumping. Thus, whenever $S_1$ is driven into absorption, $S_2$ is driven into inversion, and vice versa. From Eq.~(\ref{eq:key}) we then obtain a relation between the anomalous optical depths,
\begin{equation}
\frac{\tau_{\Ss}^{\rm an}(v)}{\mut_{\Ss}^{2}}
=-\frac{\tau_{\Sss}^{\rm an}(v)}{\mut_{\Sss}^{2}} .
\label{eq:conjratio}
\end{equation}

If the difference between the reduced transition dipole moments is neglected, the anomalous parts satisfy $\tau_{\Ss}^{\rm an}(v)=-\tau_{\Sss}^{\rm an}(v)$, the well-known relation for OH conjugate lines. Real gas, however, is never purely anomalous. The full population difference of each satellite line is the sum of an anomalous part and a `thermal' part, so the exact optical depths of conjugate satellite lines are likewise the sum of an anomalous and a `thermal' part, as shown in Fig.~\ref{fig:conjugate}b, where the two thermal components are drawn as identical curves for clarity. The optical depth identity of Eq.~(\ref{eq:rule}) then becomes
\begin{equation}
\frac{\tau_{\Ss}^{\rm ther}(v)+\tau_{\Ss}^{\rm an}(v)}{\mut_{\Ss}^{2}}
+\frac{\tau_{\Sss}^{\rm ther}(v)+\tau_{\Sss}^{\rm an}(v)}{\mut_{\Sss}^{2}}
=\frac{\tau_{M_1}^{\rm ther}(v)}{5\,\mut_{M_1}^{2}}
+\frac{\tau_{M_2}^{\rm ther}(v)}{9\,\mut_{M_2}^{2}} .
\label{eq:rule_pump}
\end{equation}
When the thermal part is weak, the satellite optical depths therefore cancel; otherwise their sum does not vanish.

A key question is how the optical depth of conjugate satellite lines is measured. From radiative transfer, the observed line brightness temperature is
\begin{equation}
T_{\rm L} = \left[\,J_{\nu}(T_{{\rm ex}})-T_{\rm bg}\,\right](1-e^{-\tau}),
\label{eq:obs_tb}
\end{equation}
where $J_{\nu}(T)=(h\nu/k)/[{\rm e}^{h\nu/(kT)}-1]$ is the Planck function expressed in brightness-temperature units, and $T_{\rm bg}$ is the brightness temperature of the background continuum source, which can be measured directly from the observations.  To obtain the optical depth, $T_{{\rm ex}}$ must be known.

\begin{figure}[htbp]
\centering
\includegraphics[width=\linewidth]{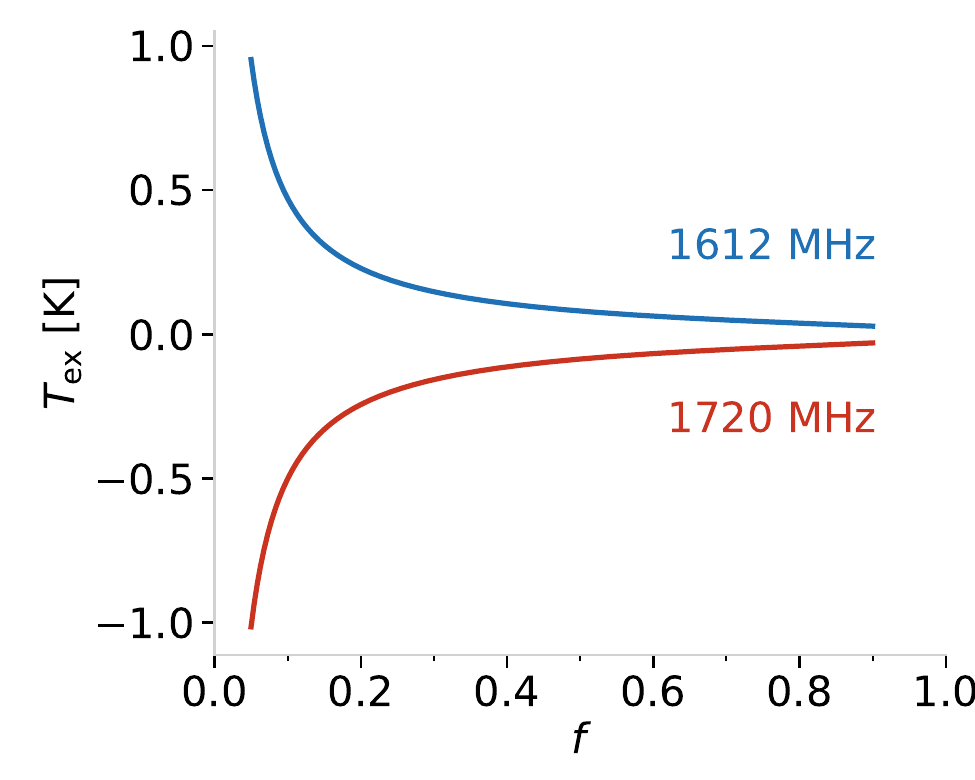}
\caption{The excitation temperatures of the OH 18 cm satellite lines as a function of $f$, the fraction of the $F=1$ population moved into $F=2$ by pumping.}
\label{fig:oh_sate_tex}
\end{figure}

For gas in LTE, the level populations satisfy
\begin{equation}
\frac{n_{ i}}{n_{ j}}=\frac{g_{ i}}{g_{ j}}{\rm exp}\left(-\frac{h\nu}{kT_{\rm kin}}\right),
\label{eq:popu}
\end{equation}
where $T_{\rm kin}$ is the gas kinetic temperature, and where the excitation temperature $T_{\rm ex}$ may replace it when the gas is non-thermal. Before pumping starts, and since $kT_{\rm kin}\gg h\nu$ is easily satisfied, we may set $n_1^-$, $n_2^-$, $n_1^+$ and $n_2^+$ to $3N$, $5N$, $3N$ and $5N$, respectively. If pumping then moves a fraction $f$ of the $F=1$ population into $F=2$, these four populations become $3N(1-f)$, $5N+3Nf$, $3N(1-f)$ and $5N+3Nf$. Equation~(\ref{eq:popu}) then gives the excitation temperatures of the satellite lines.  Fig.~\ref{fig:oh_sate_tex} shows $T_{\rm ex}$ of the OH 1612 and 1720~MHz lines as a function of $f$. There $T_{\rm ex}$ is generally small and therefore negligible compared with $T_{\rm bg}$. Consider next $J_{\nu}(T)$: when $kT_{\rm ex}\gg h\nu$ it equals $T_{\rm ex}$, and otherwise it is smaller than $T_{\rm ex}$, so in either case it is negligible compared with $T_{\rm bg}$. Consequently, the optical depths of the satellite lines can therefore be obtained from Eq.~(\ref{eq:obs_tb}). Note that the optical depth is negative when a line is in emission.

\section{Prospect}
\label{sec:prosp}
The sum rule of Eq.~(\ref{eq:rule}) is exact and free of any model for the level populations. It does not rely on the shape of a line at all, comparing the four lines channel by channel in velocity space. Because the coefficients $a$ and $b$ depend on nothing but the rotational quantum number $J$, the sum rule takes the same form for the 18~cm ground-state quartet and for the four higher-frequency quartets at 6.0, 7.8, 8.1 and 13.4~GHz. With its large collecting area and wide frequency coverage, the Square Kilometre Array (SKA) will record OH spectra at signal-to-noise ratios far beyond those of present-day instruments, and in numbers large enough to turn the search for a drift of the constants into a statistical exercise. Yet the very sensitivity that makes such data valuable also exposes the blends and the non-Gaussian wings that make it impractical to decompose each spectrum into Gaussian components and to match those components across the four transitions; the more informative the spectrum, the less the conventional analysis can be trusted. The sum rule is immune to this difficulty, so that better data translate more directly into a tighter measurement rather than into a harder analysis. Even more importantly, because the relation is applied channel by channel, channels that are contaminated by radio-frequency interference (RFI) can simply be discarded, provided that enough clean channels remain to constrain the relation \citep{su2026a}. The radio sky will only become more polluted in the SKA era, as growing satellite constellations fill it with emission, which makes the sum rules more valuable. Applied to the forthcoming SKA surveys, and equally to the large archives of OH spectra already in hand, the sum rule offers a timely and, we suggest, natural route to a precision test of whether the constants of nature are truly constant \citep{su2026a}. It is a technique whose power grows with the quality of the spectra to which it is applied, and it is therefore well matched to the era of radio astronomy that the SKA is now opening.

\section{Summary}
\label{sec:summary}

We have derived the exact relation between the optical depths of the OH $\Lambda$-doublet radio lines. The weighted population differences of the four levels obey one exact linear identity, Eq.~(\ref{eq:identity}), which can be recast into the sum rule of Eq.~(\ref{eq:rule}), with line-specific dipole moments derived from the catalogued Einstein coefficients (Table~\ref{tab:bands}). If the difference between the reduced transition dipole moments is neglected, it reduces to the sum rule reported by SG26, namely $\tau_{1612}(v)+\tau_{1720}(v)=\tau_{1665}(v)/5+\tau_{1667}(v)/9$ for the 18~cm ground state. Note that there is no sum rule for the OH lines at 4.7 GHz because the $^{2}\Pi_{1/2},J{=}1/2,F{=}0{\leftrightarrow}0$ line is strictly forbidden. Our analysis further shows that, when conjugacy occurs, the optical depths of OH 18 cm satellite lines do not always cancel out, which is natural when the thermal part is not weak.

It is worth stressing that no excitation temperature, no Rayleigh--Jeans approximation $kT\gg h\nu$, and no LTE assumption enters the derivation of the sum rule of Eq.~(\ref{eq:rule}). Consequently it applies in the same way to absorbing gas, thermally emitting gas, masers and conjugate gas. This is an advance over the results of SG26, whose derivation of the same relation relied on the Rayleigh--Jeans approximation. The OH 18 cm conjugate lines in particular may have very small excitation temperatures, which violate that approximation, so that the sum rule of SG26 cannot be applied to them. Here we have obtained the exact sum rule by a new route and have shown that it can be used more widely. As suggested by SG26, it will play a crucial role in measuring the variation of fundamental physical constants.

\section*{Acknowledgments}
We thank Yefei Yuan for reading this manuscript and providing some comments. This work uses spectroscopic data from the LAMDA database \citep{schoier2005}; for OH, the energy levels, line frequencies and Einstein A coefficients are taken from the JPL catalogue \citep{pickett1998}.

RZS acknowledges the support from the National Natural Science Foundation of China (grant 12603034), the Shanghai Super Postdoctoral Incentive Program (No. 2025322), the China Postdoctoral Science Foundation (Grant No. 2025M783232), and the National SKA Program of China (No. 2025SKA0130100). MFG is supported by the National Natural Science Foundation of China (grant 12473019), the National Science and Technology Major Project of China (No. 2024ZD1100601), the National SKA Program of China (Grant No. 2022SKA0120102), the Shanghai Pilot Program for Basic Research---Chinese Academy of Sciences, Shanghai Branch (JCYJ-SHFY-2021-013), and the China Manned Space Project (No. CMS-CSST-2025-A07).

\bibliographystyle{aasjournal}
\bibliography{oh_sum}

@ARTICLE{chengalur2003,
       author = {{Chengalur}, Jayaram N. and {Kanekar}, Nissim},
        title = "{Constraining the Variation of Fundamental Constants using 18cm OH Lines}",
      journal = {\prl},
         year = 2003,
        month = dec,
       volume = {91},
       number = {24},
          eid = {241302},
        pages = {241302},
          doi = {10.1103/PhysRevLett.91.241302},
archivePrefix = {arXiv},
       eprint = {astro-ph/0310764},
 primaryClass = {astro-ph},
       adsurl = {https://ui.adsabs.harvard.edu/abs/2003PhRvL..91x1302C}
}

@ARTICLE{combes2021,
       author = {{Combes}, F. and {Gupta}, N. and {Muller}, S. and {Balashev}, S. and {J{\'o}zsa}, G.~I.~G. and {Srianand}, R. and {Momjian}, E. and {Noterdaeme}, P. and {Kl{\"o}ckner}, H.-R. and {Baker}, A.~J. and {Boettcher}, E. and {Bosma}, A. and {Chen}, H.-W. and {Dutta}, R. and {Jagannathan}, P. and {Jose}, J. and {Knowles}, K. and {Krogager}, J.-. K. and {Kulkarni}, V.~P. and {Moodley}, K. and {Pandey}, S. and {Petitjean}, P. and {Sekhar}, S.},
        title = "{PKS 1830-211: OH and H I at z = 0.89 and the first MeerKAT UHF spectrum}",
      journal = {\aap},
         year = 2021,
        month = apr,
       volume = {648},
          eid = {A116},
        pages = {A116},
          doi = {10.1051/0004-6361/202040167},
archivePrefix = {arXiv},
       eprint = {2101.00188},
 primaryClass = {astro-ph.GA},
       adsurl = {https://ui.adsabs.harvard.edu/abs/2021A&A...648A.116C}
}

@ARTICLE{combes2023,
       author = {{Combes}, F. and {Gupta}, N. and {Muller}, S. and {Balashev}, S. and {Deka}, P.~P. and {Emig}, K.~L. and {Kl{\"o}ckner}, H.-R. and {Klutse}, D. and {Knowles}, K. and {Mohapatra}, A. and {Momjian}, E. and {Noterdaeme}, P. and {Petitjean}, P. and {Salas}, P. and {Srianand}, R. and {Wagenveld}, J.~D.},
        title = "{PKS 1413+135: OH and H I at z = 0.247 with MeerKAT}",
      journal = {\aap},
         year = 2023,
        month = mar,
       volume = {671},
          eid = {A43},
        pages = {A43},
          doi = {10.1051/0004-6361/202245482},
archivePrefix = {arXiv},
       eprint = {2211.09355},
 primaryClass = {astro-ph.GA},
       adsurl = {https://ui.adsabs.harvard.edu/abs/2023A&A...671A..43C}
}

@ARTICLE{darling2004,
       author = {{Darling}, Jeremy},
        title = "{A Laboratory for Constraining Cosmic Evolution of the Fine-Structure Constant: Conjugate 18 Centimeter OH Lines toward PKS 1413+135 at z = 0.24671}",
      journal = {\apj},
         year = 2004,
        month = sep,
       volume = {612},
       number = {1},
        pages = {58-63},
          doi = {10.1086/422450},
archivePrefix = {arXiv},
       eprint = {astro-ph/0405240},
 primaryClass = {astro-ph},
       adsurl = {https://ui.adsabs.harvard.edu/abs/2004ApJ...612...58D}
}

@ARTICLE{dawson2022,
       author = {{Dawson}, J.~R. and {Jones}, P.~A. and {Purcell}, C. and {Walsh}, A.~J. and {Breen}, S.~L. and {Brown}, C. and {Carretti}, E. and {Cunningham}, M.~R. and {Dickey}, J.~M. and {Ellingsen}, S.~P. and {Gibson}, S.~J. and {G{\'o}mez}, J.~F. and {Green}, J.~A. and {Imai}, H. and {Krishnan}, V. and {Lo}, N. and {Lowe}, V. and {Marquarding}, M. and {McClure-Griffiths}, N.~M.},
        title = "{SPLASH: the Southern Parkes Large-Area Survey in Hydroxyl - data description and release}",
      journal = {\mnras},
         year = 2022,
        month = may,
       volume = {512},
       number = {3},
        pages = {3345-3364},
          doi = {10.1093/mnras/stac636},
archivePrefix = {arXiv},
       eprint = {2203.05131},
 primaryClass = {astro-ph.GA},
       adsurl = {https://ui.adsabs.harvard.edu/abs/2022MNRAS.512.3345D}
}

@ARTICLE{dousmanis1955,
       author = {{Dousmanis}, G.~C. and {Sanders}, T.~M. and {Townes}, C.~H.},
        title = "{Microwave Spectra of the Free Radicals OH and OD}",
      journal = {Physical Review},
         year = 1955,
        month = dec,
       volume = {100},
       number = {6},
        pages = {1735-1754},
          doi = {10.1103/PhysRev.100.1735},
       adsurl = {https://ui.adsabs.harvard.edu/abs/1955PhRv..100.1735D}
}

@BOOK{elitzur1992,
       author = {{Elitzur}, Moshe},
        title = "{Astronomical masers}",
         year = 1992,
     publisher = {Kluwer},
       address = {Dordrecht},
       volume = {170},
          doi = {10.1007/978-94-011-2394-5},
       adsurl = {https://ui.adsabs.harvard.edu/abs/1992ASSL..170.....E}
}

@ARTICLE{gupta2018,
       author = {{Gupta}, N. and {Momjian}, E. and {Srianand}, R. and {Petitjean}, P. and {Noterdaeme}, P. and {Gyanchandani}, D. and {Sharma}, R. and {Kulkarni}, S.},
        title = "{Discovery of OH Absorption from a Galaxy at z {\ensuremath{\sim}} 0.05: Implications for Large Surveys with SKA Pathfinders}",
      journal = {\apjl},
         year = 2018,
        month = jun,
       volume = {860},
       number = {2},
          eid = {L22},
        pages = {L22},
          doi = {10.3847/2041-8213/aac9cd},
archivePrefix = {arXiv},
       eprint = {1806.00172},
 primaryClass = {astro-ph.GA},
       adsurl = {https://ui.adsabs.harvard.edu/abs/2018ApJ...860L..22G}
}

@ARTICLE{kanekar2005,
       author = {{Kanekar}, N. and {Carilli}, C.~L. and {Langston}, G.~I. and {Rocha}, G. and {Combes}, F. and {Subrahmanyan}, R. and {Stocke}, J.~T. and {Menten}, K.~M. and {Briggs}, F.~H. and {Wiklind}, T.},
        title = "{Constraints on Changes in Fundamental Constants from a Cosmologically Distant OH Absorber or Emitter}",
      journal = {\prl},
         year = 2005,
        month = dec,
       volume = {95},
       number = {26},
          eid = {261301},
        pages = {261301},
          doi = {10.1103/PhysRevLett.95.261301},
archivePrefix = {arXiv},
       eprint = {astro-ph/0510760},
 primaryClass = {astro-ph},
       adsurl = {https://ui.adsabs.harvard.edu/abs/2005PhRvL..95z1301K}
}

@ARTICLE{kanekar2018,
       author = {{Kanekar}, Nissim and {Ghosh}, Tapasi and {Chengalur}, Jayaram N.},
        title = "{Stringent Constraints on Fundamental Constant Evolution Using Conjugate 18 cm Satellite OH Lines}",
      journal = {\prl},
         year = 2018,
        month = feb,
       volume = {120},
       number = {6},
          eid = {061302},
        pages = {061302},
          doi = {10.1103/PhysRevLett.120.061302},
archivePrefix = {arXiv},
       eprint = {1801.07688},
 primaryClass = {astro-ph.CO},
       adsurl = {https://ui.adsabs.harvard.edu/abs/2018PhRvL.120f1302K}
}

@ARTICLE{lo2005,
       author = {{Lo}, K.~Y.},
        title = "{Mega-Masers and Galaxies}",
      journal = {\araa},
         year = 2005,
        month = sep,
       volume = {43},
       number = {1},
        pages = {625-676},
          doi = {10.1146/annurev.astro.41.011802.094927},
       adsurl = {https://ui.adsabs.harvard.edu/abs/2005ARA&A..43..625L}
}

@ARTICLE{pickett1998,
       author = {{Pickett}, H.~M. and {Poynter}, R.~L. and {Cohen}, E.~A. and {Delitsky}, M.~L. and {Pearson}, J.~C. and {M{\"u}ller}, H.~S.~P.},
        title = "{Submillimeter, millimeter and microwave spectral line catalog.}",
      journal = {\jqsrt},
         year = 1998,
        month = nov,
       volume = {60},
       number = {5},
        pages = {883-890},
          doi = {10.1016/S0022-4073(98)00091-0},
       adsurl = {https://ui.adsabs.harvard.edu/abs/1998JQSRT..60..883P}
}

@ARTICLE{robinson1967,
       author = {{Robinson}, B.~J. and {McGee}, R.~X.},
        title = "{OH Molecules in the Interstellar Medium}",
      journal = {\araa},
         year = 1967,
        month = jan,
       volume = {5},
        pages = {183},
          doi = {10.1146/annurev.aa.05.090167.001151},
       adsurl = {https://ui.adsabs.harvard.edu/abs/1967ARA&A...5..183R}
}

@ARTICLE{schoier2005,
       author = {{Sch{\"o}ier}, F.~L. and {van der Tak}, F.~F.~S. and {van Dishoeck}, E.~F. and {Black}, J.~H.},
        title = "{An atomic and molecular database for analysis of submillimetre line observations}",
      journal = {\aap},
         year = 2005,
        month = mar,
       volume = {432},
       number = {1},
        pages = {369-379},
          doi = {10.1051/0004-6361:20041729},
archivePrefix = {arXiv},
       eprint = {astro-ph/0411110},
 primaryClass = {astro-ph},
       adsurl = {https://ui.adsabs.harvard.edu/abs/2005A&A...432..369S}
}

@ARTICLE{su2025a,
       author = {{Su}, Renzhi and {An}, Tao and {Curran}, Stephen J. and {Busch}, Michael P. and {Gu}, Minfeng and {Li}, Di},
        title = "{Constraints on the Fractional Changes of the Fundamental Constants at a Look-back Time of 2.5 Myr}",
      journal = {\apjl},
         year = 2025,
        month = mar,
       volume = {981},
       number = {2},
          eid = {L25},
        pages = {L25},
          doi = {10.3847/2041-8213/adb843},
archivePrefix = {arXiv},
       eprint = {2502.14576},
 primaryClass = {astro-ph.GA},
       adsurl = {https://ui.adsabs.harvard.edu/abs/2025ApJ...981L..25S}
}

@ARTICLE{su2025b,
       author = {{Su}, Renzhi and {Curran}, Stephen J. and {Combes}, Fran{\c{c}}oise and {Gupta}, Neeraj and {Muller}, Sebastien and {Li}, Di and {Gu}, Minfeng},
        title = "{New constraints on the values of the fundamental constants at a look-back time of 7.3 Gyr}",
      journal = {\aap},
         year = 2025,
        month = jun,
       volume = {698},
          eid = {A154},
        pages = {A154},
          doi = {10.1051/0004-6361/202453407},
archivePrefix = {arXiv},
       eprint = {2505.07200},
 primaryClass = {astro-ph.CO},
       adsurl = {https://ui.adsabs.harvard.edu/abs/2025A&A...698A.154S}
}

@ARTICLE{su2026,
       author = {{Su}, Renzhi and {Curran}, Stephen J. and {Darling}, Jeremy and {Gu}, Minfeng and {Aditya}, J.~N.~H.~S. and {Tang}, Ningyu and {Li}, Di and {Zheng}, Zheng},
        title = "{A stringent constraint on the fractional change of proton g-factor}",
      journal = {Science China Physics, Mechanics, and Astronomy},
         year = 2026,
        month = feb,
       volume = {69},
       number = {4},
          eid = {249511},
        pages = {249511},
          doi = {10.1007/s11433-025-2863-7},
archivePrefix = {arXiv},
       eprint = {2511.17998},
 primaryClass = {astro-ph.CO},
       adsurl = {https://ui.adsabs.harvard.edu/abs/2026SCPMA..6949511S}
}

@ARTICLE{su2026a,
       author = {{Su}, Renzhi and {Gu}, Minfeng},
        title = "{Measuring the spatio-temporal variation of fundamental physical constants using the OH sum rules}",
      journal = {\aap},
         year = 2026,
        month = oct,
       volume = {714},
          eid = {L7},
        pages = {L7},
          doi = {10.1051/0004-6361/202661874},
archivePrefix = {arXiv},
       eprint = {2609.13962},
 primaryClass = {astro-ph.CO},
       adsurl = {https://ui.adsabs.harvard.edu/abs/2026A&A...714L...7S}
}

@ARTICLE{uzan2025,
       author = {{Uzan}, Jean-Philippe},
        title = "{Fundamental constants: from measurement to the universe, a window on gravitation and cosmology}",
      journal = {Living Reviews in Relativity},
         year = 2025,
        month = sep,
       volume = {28},
       number = {1},
          eid = {6},
        pages = {6},
          doi = {10.1007/s41114-025-00059-y},
archivePrefix = {arXiv},
       eprint = {2410.07281},
 primaryClass = {astro-ph.CO},
       adsurl = {https://ui.adsabs.harvard.edu/abs/2025LRR....28....6U}
}

@BOOK{varshalovich1988,
       author = {{Varshalovich}, D.~A. and {Moskalev}, A.~N. and {Khersonskii}, V.~K.},
        title = "{Quantum Theory of Angular Momentum}",
         year = 1988,
     publisher = {World Scientific},
       address = {Singapore},
          doi = {10.1142/0270},
       adsurl = {https://ui.adsabs.harvard.edu/abs/1988qtam.book.....V}
}

@ARTICLE{weinreb1963,
       author = {{Weinreb}, S. and {Barrett}, A.~H. and {Meeks}, M.~L. and {Henry}, J.~C.},
        title = "{Radio Observations of OH in the Interstellar Medium}",
      journal = {\nat},
         year = 1963,
        month = nov,
       volume = {200},
       number = {4909},
        pages = {829-831},
          doi = {10.1038/200829a0},
       adsurl = {https://ui.adsabs.harvard.edu/abs/1963Natur.200..829W}
}

\end{document}